\documentstyle[12pt]{article}
\begin{document}
\begin{center}
{\Large Spin-Polarised Cylinders and Torsion Balances to test Einstein-Cartan Gravity?}
\vspace{1cm}
\noindent

L.C. Garcia de Andrade\footnote{Departamento de Fisica Teorica,Instituto de F\'{\i}sica , UERJ, Rua S\~{a}o
francisco Xavier 524, Rio de Janeiro,CEP:20550-013, Brasil.e-mail:garcia@dft.if.uerj.br.}
\end{center}
\vspace{2cm}
\begin{center}
{\Large Abstract}
\end{center}
\vspace{0.5cm}
Spin-Polarised cylinders with and without axial magnetic fields are obtained as particular families of solutions of Einstein-Cartan gravity (EC).The first solution represents a spin-polarised cylinder in teleparallel gravity.The second solution is a magnetized solution representing a spin-polarised cylinder where the magnetic fields and spins are distributed along the infinite axis of the cylinder.Altough it seems that the first solution is less realist than the second it could be obtained by shielding the magnetic fields with a superconductor.The second solution is computed by taking into account the Ritter et al. experiment with the test spin-polarized mass to test spin dependent forces.Ritter experiment deals with a test mass with $>10^{23}$ spin polarized electrons which leads to a spin density of $10^{-4}gcm^{-1}s^{-1}$.  
\newpage
\section{Introduction}
Early solutions of representing spin-polarised cylinders in EC theory of gravity have been presented in the literature.Just to mention two interesting solutions we observe the Bedran-Garcia de Andrade \cite{1} cylinder where the spins are circularly polarized around the symmetry axis.The second solution by Soleng \cite{2} representing a spin-polarised solution where the spins are polarised along the infinite axis \cite{3}.In both solutions there is no sign of magnetic fields.In our solution in the second case the magnetic field is used to polarise the spins along the test mass cylinders.The first case represents a spin-polarised cylinder in Teleparallel gravity where spin-torsion density appears on the metric from the result of the Riemann-Cartan flatness which defines the teleparallelism \cite{4} as applied by Einstein in 1928
in Berlin \cite{5}.Although cylinders on the surface of the Earth leads to a Cartan torsion of about $10^{-42}$ cgs units the theoretical interest of the models discussed here remains beside the fact that as has been recently showed by Garcia de Andrade \cite{6} COBE data and G\"{o}del perturbations is possibel to obtain spin densities in the Universe as large as large as $10^{56} g cm^{-1}s^{-1}$.Also recently L\"{a}mmerzahl \cite{7} has shown that by using the splitting of spectral lines of atoms like $He^{3}$ and $Ni^{21}$ one may obtain from the Pauli spin equation in Riemann-Cartan spaces that torsion of the order $10^{-17}cm^{-1}$ may be found in the Laboratory.The Einstein-de Haas effect could also be consider with the type of geometries discussed here.In the next section we investigate the teleparallel geometry of the spin-polarised cylinder without magnetic fields.Section $3$ the spin-polarised cylinder in EC gravity with magnetic fields are considered.In this case we could try to put a rotating frame to reduce the metric obtained to the flat Minkowski metric but this is not possible because the metric coefficient would give a non-homogeneous rotation \cite{8}.
\section{Teleparallel geometry of a Spin-Polarised Cylinder}
We shall be concerned here with the Soleng \cite{2} geometry of cylinder in EC, which is given by
\begin{equation}
ds^{2}=-(e^{\alpha}dt+Md{\phi})^{2}+r^{2}e^{-{\alpha}}d{\phi}^{2}+e^{2{\beta}-2{\alpha}}(dr^{2}+dz^{2})
\label{1}
\end{equation}
where units considered have $8{\pi}G=c=1$.The functions ${\alpha},M$ and ${\beta}$ depend only on the $r$ coordinate.Orthonormal comoving tetrad frame is defined by the basis form 
\begin{equation} 
{\omega}^{0}=e^{\alpha}dt+Md{\phi}  
\label{2} 
\end{equation}
\begin{equation}
{\omega}^{1}=e^{{\beta}-{\alpha}}dr
\label{3}
\end{equation}
\begin{equation}
{\omega}^{2}=re^{-{\alpha}}d{\phi}
\label{4}
\end{equation}
\begin{equation}
{\omega}^{3}=e^{{\beta}-{\alpha}}dz
\label{5}
\end{equation}
A polarised spin density along the axis of symmetry is considered and the Cartan torsion is given in terms of differential forms as
\begin{equation}
T^{i}=2{\sigma}{\delta}^{i}_{0}{\omega}^{1}{\wedge}{\omega}^{2}
\label{6}
\end{equation}
where ${i,j=0,1,2,3}$.For computations convenience we addopt Soleng's ansatz \cite{2}  
\begin{equation}
{\Omega}={\sigma}+\frac{1}{2r}({\alpha}'M-{M'})e^{2{\alpha}-{\beta}}
\label{7}
\end{equation}
The Cartan first structure equation is  
\begin{equation}
T^{i}=d{\omega}^{i}+{\omega}^{i}_{k}{\wedge}{\omega}^{k}
\label{8}
\end{equation}
annd determines the connection forms ${\omega}^{i}_{j}$.By substitution of the connection forms into the second Cartan's structure equation
\begin{equation}
R^{i}_{j}=d{\omega}^{i}_{j}+{\omega}^{i}_{k}{\wedge}{\omega}^{k}_{j}
\label{9}
\end{equation}
we may obtain the curvature RC forms $R^{i}_{j}=R^{i}_{jkl}{\omega}^{k}{\wedge}{\omega}^{l}$ where $R^{i}_{jkl}$ is the Riemann-Cartan (RC) curvature tensor given by
\begin{equation}
R_{0101}={\Omega}^{2}+[{\alpha}"+2{{\alpha}'}^{2}-{\alpha}'{\beta}']e^{2{\alpha}-2{\beta}}
\label{10}
\end{equation}
\begin{equation}
R_{0112}=[{\Omega}'+2{\alpha}'({\Omega}-{\sigma})]e^{{\alpha}-{\beta}}
\label{11}
\end{equation}
\begin{equation}
R_{0202}={\Omega}^{2}+[\frac{{\alpha}'}{r}-{({\alpha})'}^{2}]e^{2{\alpha}-2{\beta}}
\label{12}
\end{equation}
\begin{equation}
R_{0303}=[{\alpha}'{\beta}'-{{\alpha}'}^{2}]e^{2{\alpha}-2{\beta}}
\label{13}
\end{equation}
\begin{equation}
R_{0323}=-{\Omega}({\beta}'-{\alpha}')e^{{\alpha}-{\beta}}
\label{14}
\end{equation}
\begin{equation}
R_{1201}=[{\Omega}'+2{\Omega}{\alpha}']e^{{\alpha}-{\beta}}
\label{15}
\end{equation}
\begin{equation}
R_{1212}=3{\Omega}^{2}-2{\Omega}{\sigma}+({\alpha}"+\frac{{\beta}'}{r}-{\alpha}'{\beta}'+\frac{{\alpha}'}{r})e^{2{\alpha}-2{\beta}}
\label{16}
\end{equation}
\begin{equation}
R_{1313}=({\alpha}"-{\beta}")e^{2{\alpha}-2{\beta}}
\label{17}
\end{equation}
\begin{equation}
R_{2303}=-{\Omega}({\beta}'-{\alpha}')e^{{\alpha}-{\beta}}
\label{18}
\end{equation}
\begin{equation}
R_{2323}=[\frac{{\alpha}'}{r}-\frac{{\beta}'}{r}+{\alpha}'{\beta}-{{\alpha}'}^{2}]e^{2{\alpha}-2{\beta}}
\label{19}
\end{equation}
We shall addopt here the simplest teleparallel solution $R^{i}_{jkl}=0$ which is given by considering that almost all metric functions vanish according to ${\Omega},{\alpha}$ and ${\beta}$ vanish.It is easy to check that all curvature components from (\ref{10}) to (\ref{19}) vanish while the only nonvanishing component of connection form is
\begin{equation}
{\omega}^{1}_{2}=-\frac{1}{r}{\omega}^{2}
\label{20}
\end{equation}
Since ${\Omega}$ vanishes by construction then from expression (\ref{7}) one obtains that the spin ${\sigma}$ is given by
\begin{equation}
{\sigma}=\frac{1}{2r}M'
\label{21}
\end{equation}
Since in the experiments performed by Ritter et al \cite{9} the spin density is constant and he has $>10^{23}$ spin-polarised electrons by some cubic centimeters, we also consider here the spin density ${\sigma}=constant={\sigma}_{0}$ which from expression (\ref{21}) allows us to obtain the value for the metric coefficient M as
\begin{equation}
M(r)={\sigma}_{0}r^{2}+d
\label{22}
\end{equation}
where d is an integration constant that here we may consider to vanish to simplify the final form of the teleparallel geometry of the spin-polarised cylinder
\begin{equation}
ds^{2}=-(dt+{\sigma}_{0}r^{2}d{\phi})^{2}+r^{2}d{\phi}^{2}+(dr^{2}+dz^{2})
\label{23}
\end{equation}
Note that when the spin density vanishes we are left with the flat Minkowski spacetime,therefore this metric is purely affected by spin-torsio in density of spacetime.This interior solution is obtained by considering a Weyssenhoff-Frenkel spinning fluid which obeys the Frenkel condition ${\tau}^{i}_{ji}=u^{i}S_{ji}=0$ where the spin of every particle is aligned along the infinity symmetry axis-z and the fluid has the comoving frame described above.Therefore in that frame we obtain 
\begin{equation}
{\tau}^{0}_{12}=-{\sigma}_{0}
\label{24}
\end{equation}
Thus using the above equations we obtain the only nonvanishing torsion component as 
\begin{equation}
T^{0}_{12}=2{\sigma}_{0}
\label{25}
\end{equation}
Note also that from the spin density ${\sigma}_{0}=10^{-4}g.cm^{-1}s^{-1}$
for the Ritter et al. experiment \cite{9} one may notice that the metric effect would be appreciable for a torsion balance for the spin polarised electrons when the radius of the cylindrical test mass would be of the order of $r=10^{2}cm=10 m$ which would be very difficult although not impossible to build with present technology.From the cosmological point of view we may imagine that $10 m$ is a short distance on Astronomical scales.Besides at the Planck era extremely high spin density of the order of the order of \cite{10} ${\sigma}=10^{71}gcm^{-1}s^{-1}$ maybe obtained with a radius of the order of the Planck $l_{Pl}=10^{-33}$,which would lead us to a metric deviation of the flat space of the order of $10^{5}$ cgs units.In the next section we introduce the axial magnetic field along the spin-polarised cylinder axis and solve the EC gravity field equations.
\section{Magnetized Spin-Polarized Cylinder in Einstein-Cartan-Maxwell Gravity (ECM)}
Now to solve the EC equations for the spin-polarised cylinder with magnetic field along the axis of the cylinder we introduce in the field equations the following equation stress tensor for the electromagnetic field
\begin{equation}
(T^{i}_{j})_{em}=[F^{i}_{k}F^{k}_{j}-\frac{1}{2}{\delta}^{i}_{j}(E^{2}-B^{2})]
\label{26}
\end{equation}
By considering only a $B_{z}$ magnetic field component and from expression (\ref{26})and the EC field equations one obtains explicitly
\begin{equation}
{\Omega}^{2}=\frac{1}{2}{B_{z}}^{2}
\label{27}
\end{equation}
\begin{equation}
{\Omega}^{2}-2{\Omega}{\sigma}={\rho}-\frac{1}{2}{B_{z}}^{2}
\label{28}
\end{equation}
\begin{equation}
-{\Omega}^{2}+2{\Omega}{\sigma}=p_{z}+\frac{1}{2}{B_{z}}^{2}
\label{29}
\end{equation}
\begin{equation}
{\Omega}'+{\beta}'{\Omega}=0
\label{30}
\end{equation}
By summing equations (\ref{29}) and (\ref{28}) one obtains
\begin{equation}
2{\Omega}^{2}={\rho}+p_{z}
\label{31}
\end{equation}
From expression (\ref{30}) by considering that ${\beta}$ vanishes one obtains
\begin{equation}
{\Omega}'=0
\label{32}
\end{equation}
and equation (\ref{32}) yields ${\Omega}_{0}=constant$.Notice that in teleparallel case ${\Omega}$ vanishes.From expression (\ref{7}) one thus obtain the following expression
\begin{equation}
M(r)=2({\sigma}_{0}-{\Omega}_{0})\int{rdr}
\label{33}
\end{equation}
This simple integral yields $M(r)=({\sigma}_{0}-{\Omega}_{0}){r}^{2}$.Finally from expression
\begin{equation}
{\Omega}^{2}-{\sigma}_{0}{\Omega}-({\rho}-p_{z}-{B_{z}}^{2})=0
\label{34}
\end{equation}
which is a second order algebraic equation which possess the following solutions
\begin{equation}
{\Omega}_{+}=\frac{{\sigma}+\sqrt{{\sigma}^{2}-2({\rho}-p_{z}-{B_{z}}^{2})}}{2}
\label{35}
\end{equation}
\begin{equation}
{\Omega}_{-}=\frac{{\sigma}-\sqrt{{\sigma}^{2}-2({\rho}-p_{z}-{B_{z}}^{2})}}{2}
\label{36}
\end{equation} 
Note that by substituting equation (\ref{35}) into equation (\ref{33}) one obtains
\begin{equation}
M(r)=-(\sqrt{{\sigma}^{2}-2({\rho}-p_{z}-{B_{z}}^{2})}){r}^{2}
\label{37}
\end{equation}
Therefore the geometry is affected by the Magnetic field,pressure and spin density of the polarised cylinder.A similar analysis that was done for the teleparallel metric can be done here by matching the results presented here with the experimental data to check how much these experiments affect the geometry in the sense that the geometry is deviated from flatness.A more detailed account of the ideas discussed here may appear elsewhere.
\begin{flushleft}
{\large Acknowledgements}
\end{flushleft}
I would like to thank Professor Ilya Shapiro and Prof.Rudnei Ramos for helpful discussions on the subject of this paper.Financial support from CNPq. is gratefully ackowledged.

\end{document}